# Geospatial Analysis and Internet of Things in Environmental Informatics

Andreas Kamilaris[1]  and Frank Ostermann[2]

**Abstract:** Geospatial analysis offers large potential for better understanding, modelling and visualizing our natural and artificial ecosystems, using Internet of Things as a pervasive sensing infrastructure. This paper performs a review of research work based on the IoT, in which geospatial analysis has been employed in environmental informatics. Six different geospatial analysis methods have been identified, presented together with 26 relevant IoT initiatives adopting some of these techniques. Analysis is performed in relation to the type of IoT devices used, their deployment status and data transmission standards, data types employed, and reliability of measurements. This paper scratches the surface of this combination of technologies and techniques, providing indications of how IoT, together with geospatial analysis, are currently being used in the domain of environmental research.

## 1. Introduction

The Internet of Things (IoT) includes technologies and research disciplines that enable the Internet to reach out into the real world of physical objects [1]. The vision of IoT involves seamless integration of physical devices to the internet/web, by means of well-understood, accepted and used protocols, technologies and programming/description languages, for efficient human-to-machine or machine-to-machine communication [2]. Internet-enabled things are equipped with sensors, measuring the physical world and its phenomena such as temperature, humidity, radiation, electromagnetism, noise, chemicals etc. Moreover, the rise of multi-sensory mobile phones offers advanced sensing capabilities, such as measuring proximity, acceleration and location, recording audio/noise, sensing electromagnetism or capturing images and videos [3]. An important characteristic of the measurements performed by Internet-enabled things (i.e. sensors, mobile phones, cameras etc.) is the geo-location where each measurement was done. As the physical or artificial environments are generally quite complex, being characterized by a wide variety of parameters, a rich collection of measurements in space and time is necessary in order to model and understand these ecosystems [4]. Hence, sensory-based IoT information needs to be aggregated, stored and analyzed for more elaborate and holistic reasoning and inference. To exploit this variety of IoT measurements in type, space and time, geospatial analysis is used in order to provide high-quality analytics and insights [5].

The contribution of this paper is to survey IoT applications that employ analytic operations that focus on location, examining the various geospatial analysis techniques employed. It is noted that we refer to the IoT as used in environmental informatics, i.e. sensing equipment and measurement systems recording environmental quantities. For a more general survey on IoT and geospatial analysis, the reader is encouraged to read the journal paper of [6].

---

[1] Department of Computer Science, University of Twente, The Netherlands. Email: a.kamilaris@utwente.nl
[2] Department of Geo-information Processing, University of Twente, The Netherlands. Email: f.o.ostermann@utwente.nl

## 2. Methodology

A keyword-based search for related work was performed through the web scientific indexing services *Web of Science* and *Google Scholar*, as well as in the Google search engine, using the query:

"Geospatial Analysis" AND "Internet of Things" AND "Environmental Informatics"

As this query returned only a small set of papers, we used combination of the above keywords too, as well as keywords of specific types of geospatial analysis (see Section 3) and different research areas of environmental informatics (see Table 1).

Through this process, 63 papers were initially identified. Each paper, application, case study or initiative was checked for relevance based on its abstract (or summary for non-scientific papers). From the 63 papers initially identified, 26 of them were considered highly relevant, i.e. belonged to the IoT research area and performed some meaningful geospatial analysis in the domain of environmental informatics. By belonging to IoT, we refer to the connection of the physical devices to the Internet/Web. We argue that this would not be enough for the complete IoT vision and its overall concepts, described elaborately in [7], and this is a sensitive topic discussed by [8], since many researchers only "*claim that their products follow the IoT*" (and Web of Things, WoT) specifications. In this survey work we do not deal with this aspect though, considering any research work that claims to be IoT-based.

The following research questions were considered:

1. Which geospatial analytical methods do IoT-based projects and research works use?
2. What kind of IoT device types have been used?
3. Are the IoT sensors used disposable or long-term ones? Are the IoT sensors used static or mobile?
4. Which are the IoT data transmission standards used?
5. Which is the reliability of measurements? Is some calibration of the sensors needed? How are faulty readings detected?

## 3. Analysis

This section describes the findings of our survey, based on the research questions defined previously.

Six different analytical methods of geospatial analysis and geographical information science have been recorded as the ones being used in IoT research. These methods have been derived from the *Analytical Methods* categorization as performed in the *Geographic Information Science and Technology Body of Knowledge* [9]. For each method, various IoT-based projects and initiatives using this technique have been identified. Table 1 maps each IoT research area, together with the geospatial analytical method used, according to the surveyed work of this paper. As Table 1 shows, basic analytical methods and surface analysis were the ones mostly used.

| IoT Area | Geometric Measures | Data Mining | Basic Analytical Operations | Basic Analytical Methods | Network Analysis | Surface Analysis & Geostatistics |
|---|---|---|---|---|---|---|
| Disaster monitoring | x | - | x | - | - | x |
| Wildlife monitoring | - | x | - | x | x | x |
| Agriculture | x | - | - | x | x | x |
| Environmental monitoring | - | - | x | x | x | x |
| Biodiversity | - | - | - | x | x | - |

**Tab. 1**. Relationship between IoT research areas and geospatial analysis methods used.

**3.1 IoT Devices and Sensors**

A wide variety of IoT devices has been employed in the surveyed work, including RFID tags, barcodes and QR codes [3]. Mobile phones and their embedded sensors (i.e. cameras, microphones) have been used in many studies [10], especially in studies dealing with participatory sensing and crowdsourcing [11], [12], [13]. The majority of studied papers, especially those where mobile phones were used, also harnessed GPS sensing for acquiring exact user location. Moreover, ultrasonic water sensors [14], pollution sensors [10] infrared (IR) cameras [15], temperature, humidity and pressure sensors [16], geo-cubes, meteorological and hydrological sensors [17], buoy, pressure and water column heights sensors [18], as well as 3D accelerometer and gyroscope sensors and animal collar tags [19], [20] have been recorded.

Satellites as imaging sensing devices have been harnessed in problems related to biodiversity [21], [22], livestock agriculture and environmental impact [23], [24], forest fire risk assessment [25], vulnerability assessment [26] and disaster relief [13] in earthquakes, for identifying groundwater potential zones [27] and in tsunami evacuation planning [28] (LiDAR imagery).

Furthermore, meteorological stations (i.e. air, temperature and precipitation) were used for climatological modelling [29], while radar data was also used for recording precipitation in flood forecasting [30]. In studies relevant to analysis of surfaces and geo-statistics, groundwater sample sensors and spectrophotometers for detection of heavy metals [31], surface soil sampling [32], sampling for zinc contamination [33], acoustic data, conductivity, temperature, and depth sensors [34] have been recorded.

### 3.2 Disposable Vs. Long-Term IoT Sensors

Not many papers revealed information whether the IoT devices were placed for long-term or only temporarily in order to get some measurements or perform sampling. Long-term sensor deployment was observed in water monitoring systems [14], in air quality monitoring [16], in monitoring landslide displacements [17] (using low-power 10-Watt solar panels), in deep-ocean tsunami measuring [18] (battery-powered with four-year lifetime), in weather monitoring stations [29] and in wildlife monitoring [20]. Long-term deployments were provisioned where sensor replacement was expensive or difficult.

### 3.3 Mobile Vs. Static IoT Devices

Another interesting aspect was whether the IoT sensors were used statically or they were mobile, moving in space capturing different spatial measurements in different times. Apparently, mobile phones and their sensing capabilities were used as mobile sensors [3], such as crowdsensing-based applications [11], [12], [13]. Collar tags placed on animals can be considered mobile devices [19], [20]. On the other hand, sensors were placed statically in cases of monitoring some phenomenon or potential disaster, such as landslides [17], flooding [14], [30], tsunami [18], [28], forest fire [25] and earthquakes [26]. Static sensors were also used to measure air quality [16]. Static sensors were employed in all papers related to surface analysis (i.e. interpolation) and geo-statistics (i.e. kriging), where samples were taken from specific sites, creating unified raster zones by estimating unknown values between the sampling.

### 3.4 IoT Data Transmission Standards

From the IoT projects where mobile phones were employed, Wi-Fi [10] and/or telecommunication protocols were used, such as 3G/4G and GPRS/UMTS, or even SMS/MMS messages [13]. Bluetooth was also used, together with mobile phones [3], [12].

Custom-made IoT sensors made use of the IEEE802.11 wireless transmission standard, such as ultrasonic water sensors [14] or a combination of GPRS/Wi-Fi at an air quality sensory system [16]. Moreover, a low power wide area network was observed in [19], for recording animal activity and in [15], for video transmission from IR cameras. In particular for wildlife animal monitoring [20], a combination of Bluetooth low energy (BLE) and LoRa was proposed. For landslide displacements monitoring [17], geo-cube sensors were selected, measuring meteorological and hydrological parameters based on a 2.4GHz wireless protocol. Satellite communications were involved in deep-ocean tsunami measuring [18]. Some papers did not involve any wired/wireless transmissions, recording the sensory measurements at the measuring spot, and this was the case in the majority of the papers dealing with surface analysis and geo-statistics, such as [29], [31], [32], [33].

### 3.5 Data Sources and Types

Besides geo-located information, a variety of different data sources. were used for enriching the geospatial analysis performed, such as mobile phone-based crowdsourcing [12], or satellite-based imagery [23], [25], [28], [21]. Data acquired from previous projects was used for recording soil fertility data [35]. Offline data was also used, such as governmental data repositories [24], [22], topographic sheets [31] and surveys [34].

Map layers prepared in previous research works were harnessed as well, including digitized vector maps [27], digital thematic maps [26], digital soil maps [24], and digital elevation models and maps [29], [22]. The nature of data was mostly text (i.e. data measurements from IoT sensors), images (i.e. from mobile phone cameras, satellites, spectrophotometers and fixed cameras), sound (i.e. from mobile phone microphones), or video (i.e. from mobile phone cameras and fixed cameras in farms for animal monitoring). For the case of text-based data, different data formats such as JSON, XML and CSV had been used [16].

**3.6 Reliability**

The reliability of IoT devices and the accuracy of their measurements is an important topic, however most authors did not provide any relevant details. The need for trust in participatory sensing is underlined in [11]. In crowdsourcing, sometimes the data is incomplete, such as at a noise map application [12], where the source of noise (e.g. train, airplane) was missing. Lack of complete and accurate crowdsourcing-based information complicated rescue and recovery efforts during the Haitian earthquake [13].

Lack of accuracy was a dominant issue in interpolation/kriging scenarios, where missing data created certain percentages of errors when averaging/interpolating [32]. Uncertainty in soil heavy metal pollution assessment was observed. The need to meet data quality objectives (DQOs) was stressed in [16], during the design of an air quality platform.

Dealing with missing data was an issue also in [30], towards precise flood forecasting for the river Meuse. Calibration of the sensors is important in many cases, and it was a necessary procedure in the pollution sensors used in [10]. The authors needed to send the sensors back to the manufacturer for accurately calibrating them. The problem with faulty readings is that it is sometimes hard to detect, if ground-truth information is not available to compare with. Another general problem with calibration is that *extreme values are more difficult to predict than mean values* [29]. Elaborate calibration of the high-quality tsunami measuring equipment took place in [28].

Some reliability problems occurred during installation of sensors at the field. The geo-cubes used to monitor landslide displacements in [17], experienced loss of observations due to a lack of sealing and not proper solar panels used as energy sources. The sensors used to assess irrigation water well suitability were forced to become sealed after sensing [31], because exposure to air affected the measurements. Finally, accuracy issues may appear in remote sensing, due to the resolution of satellite-based imagery used [21].

**4.     Conclusion**

This paper performed a review of research papers based on the Internet of Things, in which geospatial analysis has been employed in environmental informatics. Six different geospatial analysis methods have been identified, presented together with 26 relevant IoT initiatives adopting some of these techniques. Analysis is performed in relation to the type of IoT devices used, their deployment status and data transmission standards, data types employed, and reliability of measurements.

In general, geospatial analysis offers large potential for better understanding, modelling and visualizing our natural and artificial ecosystems, using IoT as a pervasive sensing infrastructure. This paper scratches the surface of this potential, providing some indications of how IoT, together with geospatial analysis, are currently being used in the domain of environmental research.